\documentclass[twocolumn]{aastex631}

\newcommand{\sjfull}{\textit{Swift}~J1727.8-1613}
\newcommand{\sjshort}{J1727}

\usepackage{gensymb}

\begin{document}

\title{\sjfull\, has the Largest Resolved Continuous Jet Ever Seen in an X-ray Binary}

\correspondingauthor{Callan M. Wood}
\email{callan.wood@icrar.org}

\author[0000-0002-2758-0864]{Callan M. Wood}
% \altaffiliation{Forrest Research Foundation Scholar}
\affiliation{International Centre for Radio Astronomy Research, Curtin University, GPO Box U1987, Perth, WA 6845, Australia}

\author[0000-0003-3124-2814]{James C. A. Miller-Jones}
\affiliation{International Centre for Radio Astronomy Research, Curtin University, GPO Box U1987, Perth, WA 6845, Australia}

\author[0000-0003-2506-6041]{Arash Bahramian}
\affiliation{International Centre for Radio Astronomy Research, Curtin University, GPO Box U1987, Perth, WA 6845, Australia}

\author[0000-0002-8195-7562]{Steven J. Tingay}
\affiliation{International Centre for Radio Astronomy Research, Curtin University, GPO Box U1987, Perth, WA 6845, Australia}

\author[0000-0003-3165-6785]{Steve Prabu}
\affiliation{International Centre for Radio Astronomy Research, Curtin University, GPO Box U1987, Perth, WA 6845, Australia}

\author[0000-0002-7930-2276]{Thomas D. Russell}
\affiliation{INAF, Istituto di Astrofisica Spaziale e Fisica Cosmica, Via U. La Malfa 153, I-90146 Palermo, Italy}

\author[0000-0001-8125-5619]{Pikky Atri}
\affiliation{ASTRON, Netherlands Institute for Radio Astronomy, Oude Hoogeveensedĳk 4, 7991 PD Dwingeloo, The Netherlands}

\author[0000-0002-0426-3276]{Francesco Carotenuto}
\affiliation{Astrophysics, Department of Physics, University of Oxford, Keble Road, Oxford, OX1 3RH, UK}

\author[0000-0002-3422-0074]{Diego Altamirano}
\affiliation{School of Physics and Astronomy, University of Southampton, University Road, Southampton SO17 1BJ, UK}

\author[0000-0002-6154-5843]{Sara E. Motta}
\affiliation{INAF - Osservatorio Astronomico di Brera, Via E. Bianchi 46, I-23807 Merate, Italy}

% LBA Observers
\author[0000-0002-4783-6679]{Lucas Hyland}
\affiliation{Mathematics \& Physics, School of Natural Sciences, University of Tasmania, Private Bag 37, Hobart, Tasmania, 7001}

\author[0000-0002-8978-0626]{Cormac Reynolds}
\affiliation{CSIRO Astronomy and Space Science, P.O. Box 1130, Bentley, WA 6102, Australia}

\author{Stuart Weston}
\affiliation{Space Operations New Zealand Ltd, Hargest House, PO Box 1306, Invercargill 9840, New Zealand}

% alphabetical others
\author{Rob Fender}
\affiliation{Astrophysics, Department of Physics, University of Oxford, Keble Road, Oxford, OX1 3RH, UK}

\author{Elmar K\"{o}rding}
\affiliation{Department of Astrophysics/IMAPP, Radboud University, P.O. Box 9010, 6500 GL Nijmegen, The Netherlands}

\author[0000-0003-1897-6872]{Dipankar Maitra}
\affiliation{Department of Physics and Astronomy, Wheaton College, Norton, MA 02766, USA}

\author[0000-0001-9564-0876]{Sera Markoff}
\affiliation{Anton Pannekoek Institute for Astronomy, University of Amsterdam, Science Park 904, 1098 XH Amsterdam, The Netherlands}
\affiliation{Gravitation and Astroparticle Physics Amsterdam Institute, University of Amsterdam, Science Park 904, 1098 XH 195 196 Amsterdam, The Netherlands}

\author{Simone Migliari}
\affiliation{Aurora Technology, Calle Principe de Vergara, 211, 1-B, E-28002 Madrid, Spain}

\author[0000-0002-3500-631X]{David M. Russell}
\affiliation{Center for Astrophysics and Space Science (CASS), New York University Abu Dhabi, P.O. Box 129188, Abu Dhabi, UAE}

\author[0000-0003-0167-0981]{Craig L. Sarazin}
\affiliation{Department of Astronomy, University of Virginia, 530 McCormick Road, Charlottesville, VA 22904-4325, USA}

\author[0000-0001-6682-916X]{Gregory R. Sivakoff}
\affiliation{Department of Physics, University of Alberta, CCIS 4-181, Edmonton AB T6G 2E1, Canada}

\author[0000-0002-4622-796X]{Roberto Soria}
\affiliation{INAF - Osservatorio Astrofisico di Torino, Strada Osservatorio 20, 10025 Pino Torinese, Italy}
\affiliation{College of Astronomy and Space Sciences, University of the Chinese Academy of Sciences, Beijing 100049, People's Republic of China}
\affiliation{Sydney Institute for Astronomy, School of Physics A28, The University of Sydney, Sydney, NSW 2006, Australia}

\author[0000-0003-3906-4354]{Alexandra J. Tetarenko}
\affiliation{Department of Physics and Astronomy, University of Lethbridge, Lethbridge, Alberta, T1K 3M4, Canada}

\author{Valeriu Tudose}
\affiliation{Institute for Space Sciences, National Institute for Laser, Plasma and Radiation Physics, Atomistilor 409, PO Box MG-23, 077125 Bucharest-Magurele, Romania}

% \author{others}
% \affiliation{American Astronomical Society \\
% 1667 K Street NW, Suite 800 \\
% Washington, DC 20006, USA}

% \collaboration{20}{(AAS Journals Data Editors)}

% \author{F.X Timmes}
% \affiliation{Arizona State University}
% \affiliation{AAS Journals Associate Editor-in-Chief}

% \author{Amy Hendrickson}
% \altaffiliation{AASTeX v6+ programmer}
% \affiliation{TeXnology Inc.}

% \author{Julie Steffen}
% \affiliation{AAS Director of Publishing}
% \affiliation{American Astronomical Society \\
% 1667 K Street NW, Suite 800 \\
% Washington, DC 20006, USA}


\begin{abstract}
    Multi-wavelength polarimetry and radio observations of \sjfull\, at the beginning of its recent 2023 outburst suggested the presence of a bright compact jet aligned in the north-south direction, which could not be confirmed without high angular resolution images. Using the Very Long Baseline Array and the Long Baseline Array, we imaged \sjfull\, during the hard/hard-intermediate state, revealing a bright core and a large, two-sided, asymmetrical, resolved jet. The jet extends in the north-south direction, at a position angle of $-0.60\pm0.07\degree$ East of North. At 8.4 GHz, the entire resolved jet structure is $\sim110 (d/2.7\,\text{kpc})/\sin i$ AU long, with the southern approaching jet extending $\sim80 (d/2.7\,\text{kpc})/\sin i$ AU from the core, where $d$ is the distance to the source and $i$ is the inclination of the jet axis to the line of sight. These images reveal the most resolved continuous X-ray binary jet, and possibly the most physically extended continuous X-ray binary jet ever observed. Based on the brightness ratio of the approaching and receding jets, we put a lower limit on the intrinsic jet speed of $\beta\geq0.27$ and an upper limit on the jet inclination of $i\leq74\degree$. In our first observation we also detected a rapidly fading discrete jet knot $66.89\pm0.04$ mas south of the core, with a proper motion of $0.66\pm0.05$ mas hour$^{-1}$, which we interpret as the result of a downstream internal shock or a jet-ISM interaction, as opposed to a transient relativistic jet launched at the beginning of the outburst. 
\end{abstract}

% \keywords{Classical Novae (251) --- Ultraviolet astronomy(1736) --- History of astronomy(1868) --- Interdisciplinary astronomy(804)}

\section{Introduction} \label{sec:intro}
    Black hole low-mass X-ray binaries (BH LMXBs) are ideal systems to study the launching of relativistic jets and their connection to the processes of accretion, due to their proximity and variability on human timescales. The properties of relativistic jets are strongly coupled to the properties of the accretion inflow, which both evolve dramatically and rapidly during bright outbursts. 
    
    Typical BH LMXB outbursts begin in a rising hard state, where the radio emission is dominated by a compact, steady, partially self-absorbed, synchrotron-emitting, continuous jet \citep[e.g.][]{2000A&A...359..251C,2001MNRAS.322...31F}. These hard-state jets\footnote{Throughout this letter, we refer to these continuous jets as `hard-state jets' due to their historical association with the hard state, although they can persist into the hard-intermediate state, but never into the soft state.} become quenched as the source moves towards the transition into the soft state, via intermediate states, where bright, discrete transient jet ejecta are often launched \citep[see e.g.][for a review of jets and X-ray binary outbursts]{2004MNRAS.355.1105F}. The hard-state jets are often called compact jets, due to their appearance as bright, compact point sources in high angular resolution observations. Resolved hard-state jets have only been observed in a few black hole X-ray binaries: the high-mass X-ray binary Cyg X-1 \citep{2001MNRAS.327.1273S}; and the low-mass X-ray binaries GRS 1915+105 \citep{2000ApJ...543..373D, 2004evn..conf..111R}; MAXI J1836-194 \citep{2015MNRAS.450.1745R}; and MAXI J1820+070 \citep{2021MNRAS.504.3862T}. 

    High angular resolution observations of resolved hard-state jets can provide independent probes of their fundamental jet properties, complementing other methods such as radio and infrared timing studies \citep[e.g.][]{2016MNRAS.460.3284K, 2018MNRAS.480.2054M, 2019MNRAS.484.2987T, 2021MNRAS.504.3862T}, astrometry and core shift measurements \citep[e.g.][]{2021Sci...371.1046M, 2023MNRAS.525.4426P}, and broad-band spectral modelling and spectral break studies \citep[e.g.][]{2001A&A...372L..25M,2002Sci...298..196C,2011A&A...529A...3C, 2011ApJ...740L..13G, 2014MNRAS.439.1390R, 2019MNRAS.482.2447P, 2020MNRAS.498.5772R, 2024ApJ...962..116E}.

    \subsection{\sjfull}
        \sjfull\,(hereafter \sjshort) was first detected on 2023 August 24 by \textit{Swift}/BAT \citep{2023GCN.34537....1P}, with follow-up X-ray observations revealing it to be a new candidate black hole low-mass X-ray binary in the hard-state at the beginning of a bright outburst \citep{2023GCN.34540....1K, 2023GCN.34544....1N, 2023ATel16205....1N, 2023ATel16206....1N, 2023GCN.34549....1O, 2023ATel16207....1O, 2023ATel16208....1C}. Optical observations suggested that \sjshort\, has a black hole primary with an early K-type companion star, with an orbital period of $\sim7.6$ hours at a distance of $d=2.7\pm0.3$ kpc. 

        Optical and near-infrared observations of \sjshort\, during the rising hard-state at the beginning of the outburst showed that the source was reddening, possibly due to the onset of a compact hard-state jet \citep{2023ATel16225....1B}. Radio observations from the Submillimetre Array (SMA), the Karl G. Jansky Very Large Array (VLA), the enhanced Multi Element Remotely Linked Interferometer (eMERLIN), and the Allen Telescope Array (ATA), showed a bright, flat-spectrum, unresolved source \citep{2023ATel16230....1V, 2023ATel16211....1M, 2023ATel16225....1B, 2023ATel16231....1W}. Polarisation observations at X-ray wavelengths with the Imaging X-ray Polarimetry Explorer (IXPE), at optical wavelengths with the 60 cm Tohoku telescope, at 230 GHz with the SMA, and at 5.5 and 9 GHz with the Australian Telescope Compact Array (ATCA) revealed \sjshort\, to be polarised across the electromagnetic spectrum, with a position angle consistent with being aligned in the north-south direction \citep{ 2023ATel16242....1D, 2023ATel16245....1K, 2023ApJ...958L..16V, 2023ATel16230....1V, 2023arXiv231105497I}. These observations suggested that \sjshort\, had a bright, compact, hard-state jet aligned in the north-south direction.

         In this letter we present four VLBI observations with the Very Long Baseline Array (VLBA) and the Long Baseline Array (LBA), which reveal the highly extended, resolved, north-south jet of \sjshort\, during the hard/hard-intermediate state, as well as an apparently disconnected discrete jet knot downstream. In Section~\ref{sec:methods} we detail our observations, calibration, and imaging procedures, and in Section~\ref{sec:results} we present our images and perform analysis on the jet, which we discuss in Section~\ref{sec:discussion}. 

\section{Observation, Calibration, and Imaging} \label{sec:methods}

\begin{figure}
        \centering
        \includegraphics[width=\linewidth]{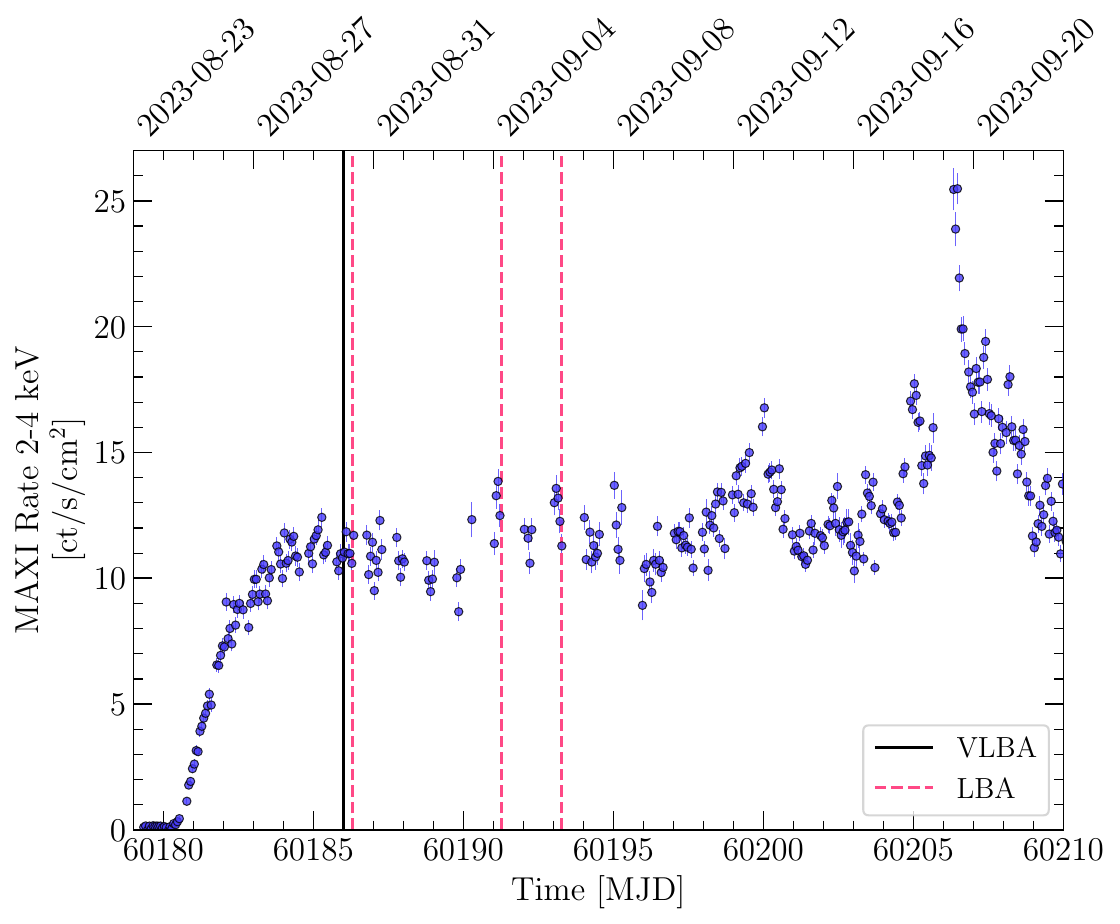}
        \caption{MAXI light curve of the evolution of \sjfull\, at the beginning of its 2023 outburst. The time of our four VLBI observations are marked by the vertical lines, and their details are given in Table~\ref{tab:vlbi observation log}.}
        \label{fig:x-ray lightcurve}
    \end{figure}

    Following the beginning of the outburst and the detection of a bright radio counterpart \citep{2023ATel16211....1M}, we observed \sjshort\, with the VLBA (project code BM538) at 8.37 GHz on 2023 August 30, as part of the Jet Acceleration and Collimation Probe Of Transient X-Ray Binaries \citep[JACPOT XRB;][]{2011IAUS..275..224M} program. Following that observation, we observed $~\sim4$ hours later with the LBA (project code V456) at 8.44 GHz, and then twice more on 2023 September 04 and September 06. Figure~\ref{fig:x-ray lightcurve} shows the timing of our observations during the beginning of the outburst as seen by the Monitor of All-sky X-ray Image \citep[MAXI\footnote{http://maxi.riken.jp/};][]{2009PASJ...61..999M} mission. The observation details can be found in Table~\ref{tab:vlbi observation log}. Further VLBI observations of the evolution of \sjshort~throughout its outburst will be presented in a future paper (Wood et al. in prep.).

    \begin{table*}
        \centering
        \caption{VLBI observation log of the \sjfull~hard-state jet.}
        \label{tab:vlbi observation log}
        \begin{tabular}{ccccccccc}
            \hline
            Label & Date    & Time  & MJD & Telescope & Observation & Frequency & Bandwidth & Stations$^\dagger$\\
            & (2023) & (UTC) & (Midpoint)  &  & Code        & (GHz)     & (MHz)    \\ \hline
            VLBA & 30~Aug   & 00:32:46--03:20:37& 60186.08    & VLBA      & BM538A     & 8.37       & 512 & FD,HN,KP,LA, \\ 
            &&&&&&&&                                                                                        MK,NL,OV,SC  \\
            LBA1 & 30~Aug   & 07:05:56--12:46:35& 60186.41    & LBA       & V456H      & 8.44       & 64 & CD,HB,KE,MP,\\
            &&&&&&&&                                                                                       PA,WW        \\
            LBA2 & 04~Sep   & 06:35:56--12:19:59& 60191.39    & LBA       & V456I      & 8.44       & 64 & AT,CD,HB,KE, \\ 
            &&&&&&&&                                                                                       MP,PA,WW     \\
            LBA3 & 06~Sep   & 06:36:02--12:19:59& 60193.39    & LBA       & V456J      & 8.44       & 64 & AT,CD,MP,PA,  \\ 
            &&&&&&&&                                                                                        WW           \\\hline
        \end{tabular}
        
        $^\dagger$ LBA Stations: CD=Ceduna, HB=Hobart 12m, KE=Katherine, MP=Mopra, PA=Parkes, WW=Warkworth 30m, AT=ATCA
    \end{table*}

    For the VLBA, we used ICRF J174358.8-035004 (J1743-0350) as a fringe finder, ICRF J172134.6-162855 (J1721-1628) as a phase reference source, and ICRF J172446.9-144359 (J1724-2914) as a check source \citep{2020A&A...644A.159C}. We observed geodetic blocks \citep{2009ApJ...700..137R} for $\sim$30 minutes at the beginning and end of the observation to improve astrometric calibration. For the LBA, we used ICRF J192451.0-291430 \citep[B1921-293;][]{2020A&A...644A.159C} as a fringe finder. In order to maximise signal-to-noise on the long baselines for phase calibration, we swapped the phase reference and check sources for the LBA. The data were correlated using the DiFX software correlator \citep{2007PASP..119..318D, 2011PASP..123..275D}, and calibrated according to the standard procedures within the Astronomical Image Processing System \citep[\textsc{aips}, version 31DEC22;][]{1985daa..conf..195W, 2003ASSL..285..109G}. After the standard external gain calibration, we performed several rounds of hybrid mapping of the phase reference source to derive the time-varying phase, delay, and rate solutions, which we interpolated to \sjshort. We also performed a single round of amplitude self-calibration to get most accurate time-varying amplitude gain calibration, which we applied to \sjshort. To match the flux density scales of the VLBA and the LBA, we used the VLBA map of J1724-2914 to derive a global amplitude gain solution for each LBA antenna IF, and polarisation to scale the a priori amplitude gains approximated from the zenith system equivalent flux densities. We used the VLBA and LBA scans of J1721-1628 to confirm that the LBA flux density scale matched the VLBA to within 5\%.

    We imaged \sjshort\, within \textsc{aips} using the CLEAN algoritm \citep{1974A&AS...15..417H} with natural weighting to maximise sensitivity, and we performed multiple rounds of phase-only self-calibration followed by a single round of amplitude self-calibration to obtain the final images. 

\section{Images and Analysis} \label{sec:results}
   \begin{figure}
        \centering
        \includegraphics[width=0.65\linewidth]{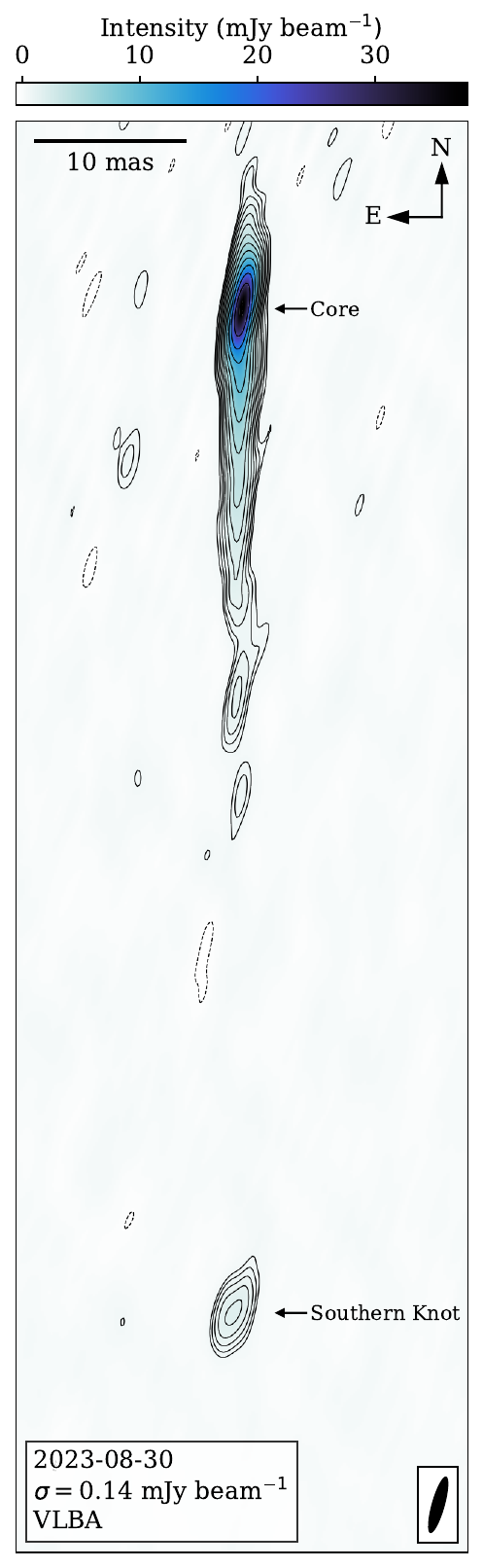}
        \caption{First VLBI image of \sjfull\, with the VLBA. The contours mark $\pm\sigma\times\sqrt{2}^n$ mJy~beam$^{-1}$ where $n=3,4,5,...$, and $\sigma$ is the rms noise shown in the lower left of the image. The ellipse in the lower right corner shows the synthesised beam. The image shows a bright core with highly resolved ($\sim40$ mas) apparently asymmetric bipolar jets extending in the north-south direction, and a discrete jet knot to the south at a separation of $66.7\pm0.2$ mas from the core at a position angle of $179.4\pm0.15$\degree\, East of North.}
        \label{fig:VLBA Image 1}
    \end{figure}
    
    Figure~\ref{fig:VLBA Image 1} shows the VLBA image of \sjshort~from 2023 August 30, which reveals a bright core with a highly resolved, asymmetric jet extending in the north-south direction, and a discrete jet knot to the south at a separation of $66.7\pm0.2$ mas from the core (which we later refined with visibility modelling). The extended continuous structure is a total of $\sim40$ mas in length, with the southern and northern jets extending $\sim30$ and $\sim10$ mas from the core, respectively. 

    We measured the position of the core in the VLBA image by fitting a point source to the brightest region of the jet using the \textsc{aips} task \texttt{JMFIT}, prior to applying any phase self-calibration. This gave a position (in the FK5 reference frame and the J2000 equinox) of
    $$
            \text{R.A.} = 17^{\text{h}} 27^{\text{m}} 43^{\text{s}}.3135784\pm0.0000065, 
            % \text{R.A.} = 17^{\text{h}} 27^{\text{m}} 43^{\text{s}}.3135784\pm0.0000003, 
    $$
    $$
            \text{Dec.} =-16\degree 12^\prime 19^{\prime\prime}.18042\pm0.00033,
            % \text{Dec.} =-16\degree 12^\prime 19^{\prime\prime}.180424\pm0.000011,
    $$
    where the errors are the $1\sigma$ statistical errors reported by AIPS added in quadrature with the estimated VLBA systematic astrometric errors \citep{2006A&A...452.1099P}, and assuming the position of J1721-1628 to be $\text{R.A.}=17^{\text{h}} 22^{\text{m}} 56^{\text{s}}.498932\pm0.000077, \text{Dec.}=-16\degree 30^\prime 19^{\prime\prime}.26363\pm0.00072$.

    By fitting a constant position angle to the locations of the positive flux density CLEAN components in the VLBA observations, we measured the position angle of the jet to be $-0.60\pm0.07\degree$ East of North. This is consistent with the position angle from the core to the southern discrete jet knot. The jet is unresolved perpendicular to the jet axis for its entire length, so based on the size of the beam and the length of the jet, we place an upper limit on the apparent jet half-opening angle of $<0.5\degree$.

\begin{figure*}
        \centering
        \includegraphics[width=\linewidth]{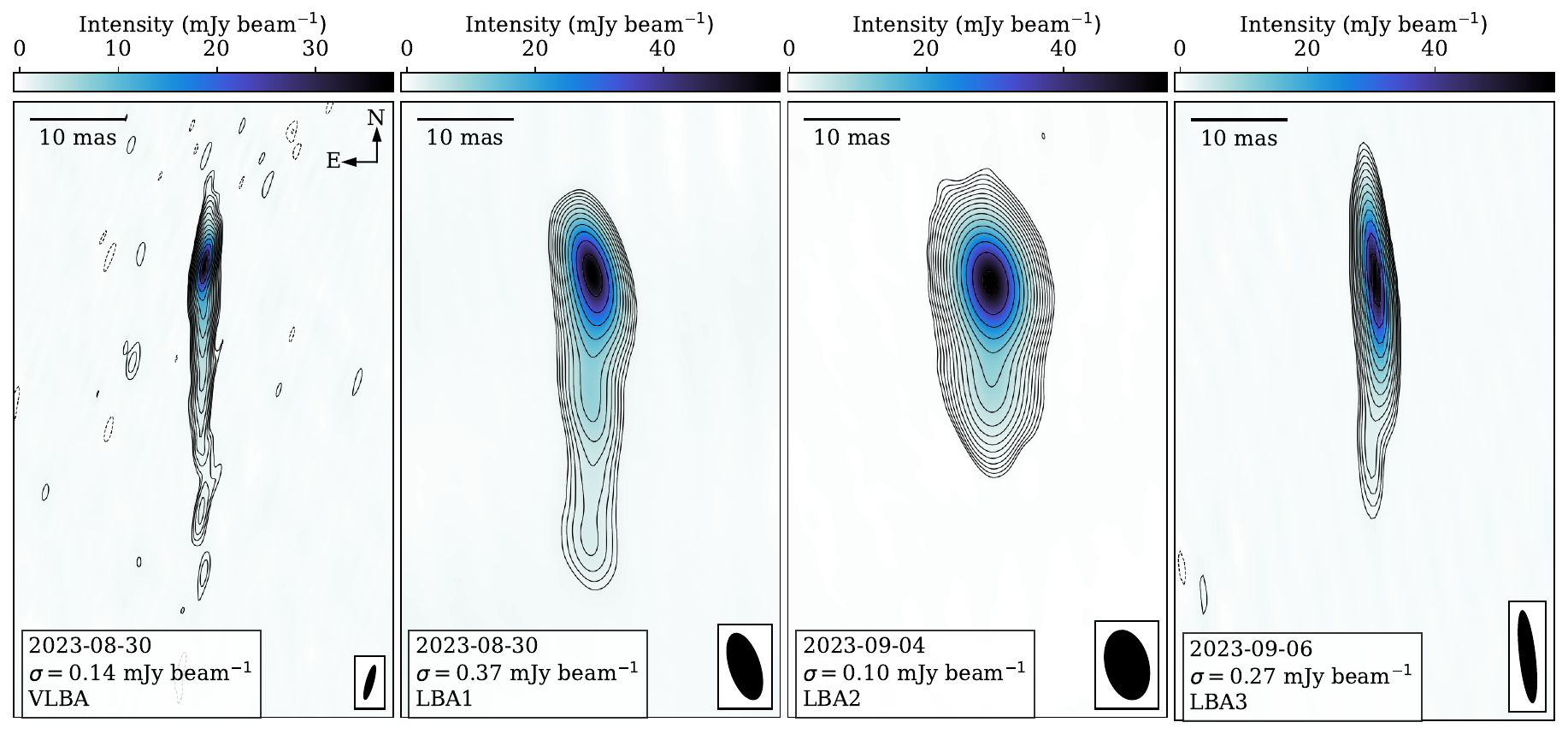}
        \caption{VLBI images of the resolved jet of \sjfull\; during the hard/hard-intermediate state. The contours mark $\pm\sigma\times\sqrt{2}^n$ mJy~beam$^{-1}$ where $n=3,4,5,...$, and $\sigma$ is the rms noise shown in the lower left of each image. The ellipse in the lower right corner of each image shows the synthesised beam. The observation parameters can be found in Table~\ref{tab:vlbi observation log} and the image parameters can be found in Table~\ref{tab:Jet parameters}. The jet is resolved over multiple epochs during the hard state, but appears less extended as the source moves towards the hard-intermediate state.}
        \label{fig:montage}
    \end{figure*}

    In Figure~\ref{fig:montage} we show all four observations of the resolved jet of \sjshort. These observations similarly show a bright core with an extended jet at the same position angle, however the LBA images have poorer angular resolution than the VLBA image, and thus we only resolve the southern jet. We do not see the southern discrete jet knot in any of the LBA observations. We summarise the observed peak and integrated flux densities of the observations in Table~\ref{tab:Jet parameters}. The integrated flux density of the jet decreased from the first observation to the final two observations. 
    
    The LBA was slightly more sensitive to diffuse emission than the VLBA, which explains why we recovered slightly more emission at the tip of the southern jet in the first LBA observation than with the VLBA. By imaging the VLBA observation with only the shortest baselines, we were unable to detect any more diffuse emission between the extended jet and the discrete jet knot. We note that the VLBA was missing the Pie Town station, and therefore lacked a few key short baselines, which could explain why we only partially detect the diffuse emission at the tip of the southern continuous jet, causing its knotty appearance. The measured flux densities from our first two observations are consistent with the observations from eMERLIN and the ATA on 2023 August 29 and August 30, respectively \citep{2023ATel16231....1W, 2023ATel16228....1B}, to within the $\sim$10\% amplitude calibration uncertainties, suggesting that we have recovered most of the emission of the jet of \sjshort\, in our observations. 

    \begin{table*}
        \centering
        \caption{Summary of the jet image parameters from Figure~\ref{fig:montage}. The core flux density is calculated by fitting a point source to the core. Here we add a 10\% amplitude calibration error in quadrature with the $1\sigma$ statistical errors reported by \textsc{aips}. The statistical error in the integrated flux density is given by $\sigma\sqrt{N_{\text{beam}}}$, where $\sigma$ is the rms noise in the image and $N_{\text{beam}}$ are the number of independent beams in the integrated area.}
        \begin{tabular}{cccccc}
            \hline
           Observation  & Integrated Flux Density & Core Flux Density & Beam Dimensions & Beam Position Angle \\
           & (mJy) & (mJy) & (mas) x (mas) & (\degree\,East of North)\\\hline
            VLBA & $101\pm10$ & $46\pm5$&3.9 x 0.9&-15.7\\
            LBA1 & $90\pm16$ & $64\pm6$&7.3 x 3.3&17.7\\
            LBA2 & $71\pm8$ & $58\pm6$&7.6 x 4.7&14.0\\
            LBA3 & $75\pm8$ & $65\pm7$&9.9 x 1.7&6.5\\            \hline
        \end{tabular}
        \label{tab:Jet parameters}
    \end{table*}

    For symmetric, continuous, steady-state jets inclined to the line of sight, the approaching and receding jets should be approximately symmetric, however the apparent surface brightness of these jets will be asymmetric due to relativistic aberration. The flux density ratio of approaching and receding continuous jets is given by
    \begin{equation}\label{eqn:flux density ratio}
        \frac{S_a}{S_r} = \left(\frac{1+\beta\cos i}{1-\beta\cos i}\right)^{2-\alpha},
    \end{equation}
    where $S_a$ and $S_r$ are the flux densities of the jets at an equal angular separation from the core, and $\alpha$ is the spectral index of the jets \citep[$S_\nu\propto\nu^{\alpha}$;][]{1967MNRAS.136..123R, 1979Natur.277..182S, 1999ARA&A..37..409M}. The emission from the receding jet will fall below the noise limit of an observation at a smaller separation than the approaching component, which explains the asymmetry seen in our images if the northern jet is receding and the southern jet is approaching.

    Since the receding jet was only resolved in the VLBA observation, we used this observation to constrain the flux density ratio. The CLEAN components for the northern jet terminated at a separation of $6.5$ mas from the core, and so we integrated the flux density of the CLEAN components along the northern and southern jets from 3-6.5 mas separation from the core. We chose the lower limit of 3 mas to avoid the emission of the unresolved core (see Appendix~\ref{appendix:jet profile}). This yielded a flux density ratio of $4.8\pm0.4$, which gave $\beta\cos i=0.29\pm0.03$, assuming a spectral index of $\alpha=-0.6\pm0.2$ (since the extended jet should be optically thin beyond the core). 

    \subsection{Modelling the Southern Knot}
    
    By separately imaging the first and second halves of the VLBA observation, we found that the southern knot appeared to be moving away from the core. We were able to precisely measure this motion by implementing the time-dependent model-fitting procedure described in \citet{2023MNRAS.522...70W}, where we fit a time-evolving model directly to the measured visibilities using the Bayesian inference algorithm nested sampling \citep{10.1214/06-BA127} implemented in the \texttt{dynesty}\footnote{\url{https://github.com/joshspeagle/dynesty}} Python package \citep{2020MNRAS.493.3132S}. Before modelling the knot, we subtracted the CLEAN components of the extended jet from the visibilities to create a residual observation containing only the southern knot. We then fit an analytical model directly to these residual visibilities, consisting of a circular Gaussian with fixed size moving away from the core at a fixed position angle ($\theta$) with a constant speed ($\dot r$). We allowed the flux density ($F$) of this component to vary linearly within the observation. The flux density evolved as
    \begin{equation}\label{eqn:flux density}
        F(t) = F_0 + \dot F (t-t_0),
    \end{equation}
    and the position of the knot relative to the core as
    \begin{equation}\label{eqn:position x}
        \Delta x(t) = (r_0 + \dot r (t-t_0))\sin\theta,
    \end{equation}
    and
    \begin{equation}\label{eqn:position y}
        \Delta y(t) = (r_0 + \dot r (t-t_0))\cos\theta,
    \end{equation}
    where $t_0$ is the reference time chosen as the approximate midpoint of the observation (02:00:00 UTC on 2023 August 30th), $r_0$ is the separation of the jet knot from the core (in mas) at the reference time, and $\Delta x$ and $\Delta y$ are in the directions of positive R.A. and Dec., respectively (in mas). We placed Gaussian priors on the separation of the jet knot at the reference time, and its position angle, based on the location of the knot in the image. We placed uniform priors on all other parameters. The prior distributions are listed in Table~\ref{tab:modelfit parameters} along with the posterior estimates. We also tried models that included linear expansion of the jet knot, but found that the expansion speed was consistent with zero. Similarly, we were unable to measure any intra-observational deceleration. We also found that models with exponential or power-law flux density decays were indistinguishable from a linear decay model on the timescale of the $\sim3$ hour observation. 

    \begin{table*}
        \centering
        \caption{Prior distributions and posterior estimates for the moving jet knot in the VLBA observation in Figure~\ref{fig:VLBA Image 1}, using equations~\ref{eqn:flux density}-\ref{eqn:position y}. The reference time is defined as 02:00:00 (UTC) on 2023 August 30th, which is approximately the midpoint of the observation. We report the median of the marginal posterior distributions as the best-fit parameters, and the 16th and 84th percentiles as the uncertainties.}
        \begin{tabular}{llr}
            \hline
             Parameter & Prior Distribution & Posterior Estimate \\\hline
             $F_0$ (mJy) & $\mathcal{U}(\text{min}=0,\text{max}=10)$& $4.60\pm0.10$\\
             $\dot{F}$ (mJy hour$^{-1}$) & $\mathcal{U}(\text{min}=-10,\text{max}=10)$& $-1.11\pm0.09$\\
             FWHM (mas) & $\mathcal{U}(\text{min}=0,\text{max}=10)$& $1.92\pm0.05$\\
             $r_0$ (mas) & $\mathcal{N}(\mu=65,\sigma=5)$& $66.89\pm0.04$\\
             $\dot{r}$ (mas hour$^{-1}$) & $\mathcal{U}(\text{min}=0,\text{max}=10)$& $0.66\pm0.05$\\
             $\theta$ (\degree\,East of North) & $\mathcal{N}(\mu=180,\sigma=5)$& $179.24\pm0.02$ \\\hline
        \end{tabular}
        
        \label{tab:modelfit parameters}
    \end{table*}

    We found that the knot was clearly resolved, and was moving away from the core at a proper motion of $\dot r = 0.66\pm0.05$ mas hour$^{-1}$, at a position angle consistent with the extended jet position angle. We found that the jet knot was rapidly decaying over the observation, which explains why we do not see the knot in the first LBA observation $\sim4$ hours later. Using the separation of the knot and its ballistic speed, we calculated an ejection date of MJD $60181.8\pm0.2$, although in Section~\ref{sec:southern jet knot} we argue that this knot is not a discrete transient jet. The separation and FWHM size of the jet knot in the VLBA observation (see Table~\ref{tab:modelfit parameters}) give a projected half-opening angle of $0.822\pm0.021$\degree.

    To ensure that the process of self-calibration was not corrupting or inducing any false variability in the data, we performed the subtraction and modelling process on versions of the observation without any self-calibration, after phase-only self-calibration, and after amplitude and phase self-calibration. We found that the position and motion of the knot was consistent between all three self-calibration scenarios, and that the size and flux density changed slightly after amplitude self-calibration. We also found marginal evidence of expansion in the observation containing amplitude self-calibration, however we only report the model parameters from the fit to the observation containing phase-only self-calibration, since we could not confirm that the amplitude self-calibration was not inducing false structure and variability in the jet knot.

\section{Discussion}\label{sec:discussion}
    We have observed the highly resolved jet of \sjfull\, over four epochs with the VLBA and the LBA. The jet is continuous, extended, apparently asymmetrical, and aligned in the north-south direction. In the first observation there was also a discrete jet knot moving away from the core to the south, which does not appear in the subsequent observations. We were able to constrain the motion and flux density variability of the jet knot with time-dependent visibility model-fitting. 

    \subsection{Core Location}

    {We measured the location of the bright core in the VLBA image, which is likely consistent with the location of the central black hole. In Cyg X-1, where the base of the jet is highly free-free absorbed by the strong stellar wind of the supergiant companion \citep[see e.g.][]{2021Sci...371.1046M}, the distance between the black hole and the synchrotron photosphere (where the compact ``core" emission originates from) is between $\sim1-3\times10^{13}$ cm at 8.4 GHz \citep{2023ApJ...951L..45Z}. Accounting for the scaling of the photosphere distance ($z_0$) with luminosity as $z_0\propto L_\nu^{\sim0.47}$ \citep{2006ApJ...636..316H, 2023ApJ...951L..45Z}, this would correspond to a distance of no more than 1-3 mas downstream from the location of the black hole in \sjshort, which is within the beam of the VLBA observation projected along the jet axis. Here we have assumed that \sjshort\, has a similar opening angle, inclination angle, and jet speed to Cyg X-1. If the inclination of the jet in \sjshort\, is larger than the normalisation in \citet{2023ApJ...951L..45Z} ($i=27.5$\degree) then this distance will be smaller. Since \sjshort\, contains an early K-type dwarf companion \citep{2024A&A...682L...1M}, the jet is not absorbed by a strong stellar wind, and thus this is a very conservative upper limit on the distance from the black hole to the jet photosphere.
    
    We also saw no evidence of a systematic shift in the position of the core as the jet faded. We note that we had to correct for a slight offset due to the different phase calibrators for the VLBA and LBA. Subsequent observations with the VLBA later in the outburst as the extended jet contracted and transient ejecta were launched showed no significant shift in the location of the core (Wood et al. in prep.). This suggests that the distance between the photosphere and the central black hole is not significant compared to the size of the restoring beam.
    
    Any core offset in \sjshort\, leads to an increased brightness ratio, which essentially increases the intrinsic jet speed at a given jet inclination (i.e. the brightness ratio curve constraint shifts right in Figure 5). Although we do not believe there is evidence for a significant core offset, here we report the changes for an offset of 1 mas. In such a case the estimated brightness ratio would increase to $7.1\pm0.8$, and $\beta\cos i = 0.36\pm0.03$, which would slightly increase the lower limit on the intrinsic jet speed and decrease the upper limit on the jet inclination (see Section~\ref{sec:jet speed}). We emphasize that such a change has no strong impact on the interpretations of this paper.

    \subsection{Resolved Jet}

    The VLBA image of \sjshort\, (Figure~\ref{fig:VLBA Image 1}) is the most resolved image of an X-ray binary hard-state jet. Cyg X-1 and GRS 1915+105 are the only other X-ray binaries with a hard-state jet that has been resolved in an image over multiple synthesised beams \citep{2001MNRAS.327.1273S, 2000ApJ...543..373D, 2004evn..conf..111R}. Assuming a distance to \sjshort\, of $2.7\pm0.3$ kpc \citep{2024A&A...682L...1M}, the size of the entire resolved jet structure (including the receding jet) was $\sim110/\sin i$ AU, and the extent of the approaching jet was $\sim80/\sin i$ AU, or $5\times10^8/\sin i\,r_g$ \citep[assuming a black hole mass of $8\,M_\odot$][]{2012ApJ...757...36K}. X-ray polarisation observations of \sjshort\, with IXPE suggested that the inclination of the inner accretion flow is between $\sim30\degree-60\degree$ \citep{2023ApJ...958L..16V}. Therefore the physical extent of the approaching resolved jet of \sjshort\, on 2023 August 30th is between $\sim95-160$ AU, or $\sim(0.6-1)\times10^9\,r_g$. 

    The two-sided jets of both GRS 1915+105 and Cyg X-1 have been resolved at 8.4 GHz \citep{2000ApJ...543..373D, 2004evn..conf..111R, 2021Sci...371.1046M}. The approaching jet of Cyg X-1 has been detected out to an extent of $\sim50$ AU \citep{2001MNRAS.327.1273S, 2021Sci...371.1046M}. In 1997, the approaching jet of GRS 1915+105 was detected out to $\sim16$ AU \citep[updated with the most recent distance constraint;][]{2000ApJ...543..373D, 2023ApJ...959...85R}, and in 2003 the jet was measured to be $\lesssim150$ AU \citep[with the upper limit due to interstellar scatter broadening]{2004evn..conf..111R}. The approaching jet of MAXI J1836-194 (which has a poorly constrained distance and inclination) was marginally resolved at 8.4 GHz during the decay of its 2011 outburst, with an extent of $\sim7-115$ AU \citep[][]{2014MNRAS.439.1390R, 2014MNRAS.439.1381R, 2015MNRAS.450.1745R}. The approaching hard-state jet of MAXI J1820+070 was marginally resolved at 15 GHz, corresponding to a physical extent of $\sim3$ AU at 8.4 GHz \citep[scaling the size as $z\propto1/\nu$;][]{1979ApJ...232...34B, 2021MNRAS.504.3862T}. 
    
    The hard-state jet of \sjshort\, was detected out to a further extent than the hard-state jets of Cyg X-1, GRS 1915+105 in its 1997 flare, and MAXI J1820+070. Depending on inclination, the resolved hard-state jet of \sjshort\, may also have been larger than the jet of GRS 1915+105 during its 2003 flare, and the decaying hard-state jet of MAXI J1836-194, and therefore \sjshort\, may have had the most extended hard-state jet ever observed in an X-ray binary. The extent of a resolved a hard-state jet is the distance from the core at which the surface brightness of the expanding jet material falls below the noise limit. This depends on both the physical properties of the jet (e.g. opening angle, jet speed, jet content, magnetic field strength, and jet internal structure), as well as the observation parameters, particularly the angular resolution, the overall sensitivity, and the sensitivity to diffuse structure. 

    Our measured position angle of both the extended jet and the jet knot is consistent with the radio, mm, and X-ray polarisation position angles of \sjshort, which we compare in Figure~\ref{fig:polar}. These measurements are also consistent with the more poorly constrained optical polarisation orientation \citep{2023ATel16245....1K}. These measurements suggest that structures in the accretion flow, the jet base, and the downstream jet are aligned to within a few degrees in the plane of the sky.

    \begin{figure}
        \centering
        \includegraphics[width=\linewidth]{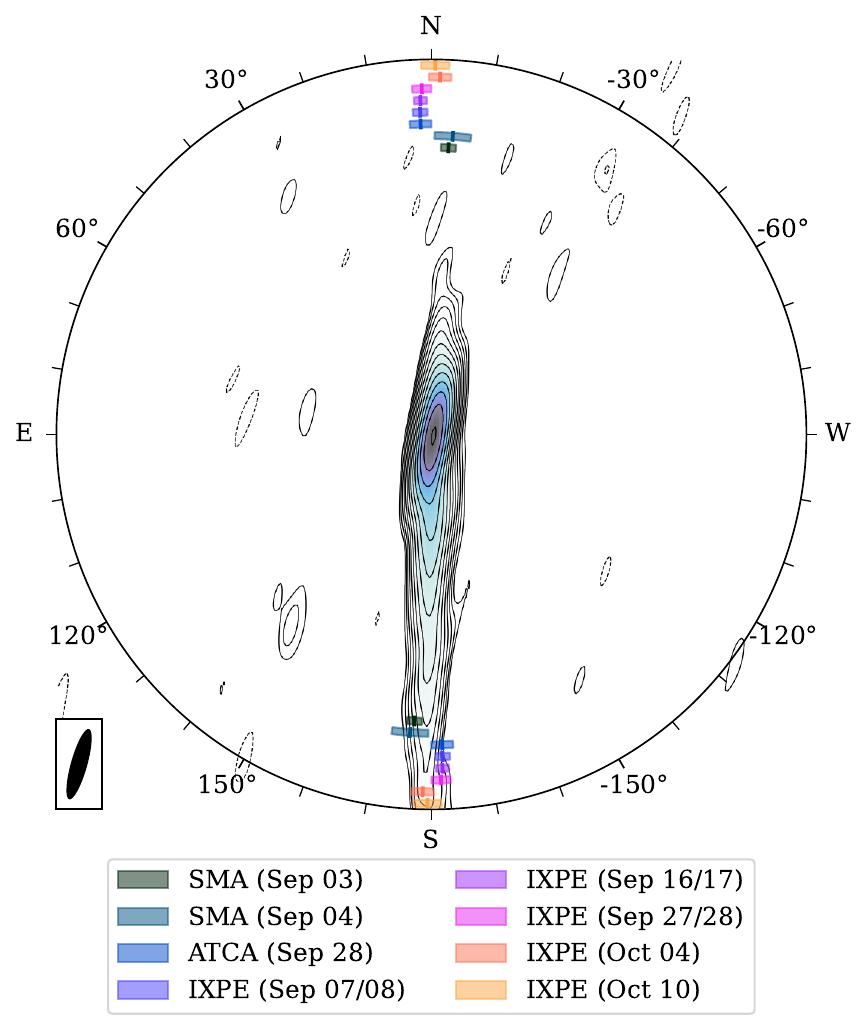}
        \caption{Comparison of polarisation angles and the position angle of the resolved jet of \sjfull. The background image is the same as shown in Figure~\ref{fig:VLBA Image 1}. We plot the polarisation measured by the Sub-Millimeter Array at 1.3mm \citep[SMA;][]{2023ATel16230....1V}, the Australia Telescope Compact Array at 5.5 and 9 GHz \citep[ATCA;][]{2023arXiv231105497I}, and the Imaging X-ray Polarimetry Explorer \citep[IXPE;][]{2023ATel16242....1D,2023ApJ...958L..16V, 2023arXiv231105497I}. The shaded arcs represent the reported $1\sigma$ error intervals, with the lines showing their nominal values. The polarisation observations span the hard and hard-intermediate states and are consistent with the north-south orientation of the resolved jet in our VLBI observations.}
        \label{fig:polar}
    \end{figure}

    Between the observations on 2023 August 30th and September 4th and 6th, the extended continuous jet became fainter and less extended, while the core remained relatively constant. We note that the noise level in the observations changes, and thus the distance from the core to which we could detect an identical jet in each observation is different. Our LBA observations did not have sufficient angular resolution and sensitivity to resolve the counter jet, and so we cannot comment on how the jet speed changed as the compact jet faded. We cannot use the lack of a detected counter jet to place a lower limit on the brightness ratio, since the northern jet in the VLBA image can be completely contained within the beam of the subsequent LBA observations. Given the changing resolution of the observations, we also cannot meaningfully compare the peak intensity of the unresolved core across the epochs.

    \subsection{Intrinsic Jet Speeds}\label{sec:jet speed}
    Without observing a northern counterpart to the southern jet knot, we cannot uniquely constrain the intrinsic speed and inclination of the source, however we can constrain possible combinations of $\beta$ and $i$. The proper motion of the approaching jet knot, $\mu$, is related to the intrinsic speed of the jet knot, $\beta$, by,
    \begin{equation}\label{eqn:proper motion}
        \mu=\frac{\beta\sin i}{1-\beta\cos i}\frac{c}{d},
    \end{equation}
    where $i$ is the inclination angle of the jet to the line of sight, $d$ is the distance to \sjshort\, \citep[$2.7\pm0.3$ kpc;][]{2024A&A...682L...1M}, and $c$ is the speed of light \citep{1999ARA&A..37..409M}. We cannot use the non-detection of a receding counterpart to set an upper limit on the flux density ratio to constrain $\beta\cos i$, since we need to measure the flux densities when the jets are at equal separations (i.e. at the same age) since the jets vary in brightness as they move downstream \citep{2004ApJ...603L..21M}. 

    In Figure~\ref{fig:jet speed inclination} we plot the possible values of $\beta$ and $i$ for both the extended jet and the jet knot, using equations~\ref{eqn:flux density ratio} and \ref{eqn:proper motion}, respectively. These constraints place a lower limit on the intrinsic jet speed of both the continuous and discrete jet knot of $\beta\geq0.27$ and $\beta\geq0.2$, respectively. The brightness ratio also gives an upper limit on the jet inclination of $i\leq74\degree$. Although they do not have to share the same intrinsic speed, assuming that the discrete jet knot and the continuous extended jet are at the same inclination \citep[i.e. assuming no rapid large-scale precession of the jet axis, as in V404 Cygni;][]{2019Natur.569..374M}, then their intrinsic speeds are consistent within $1\sigma$ in the interval $i\sim27-37\degree$. In the inclination range favoured by X-ray polarisation measurements \citep[$i\sim30\degree-60\degree$;][]{2023ApJ...958L..16V}, the jet knot is moving at the same speed or slower than the continuous jet.  

    \begin{figure}
        \centering
        \includegraphics[width=\linewidth]{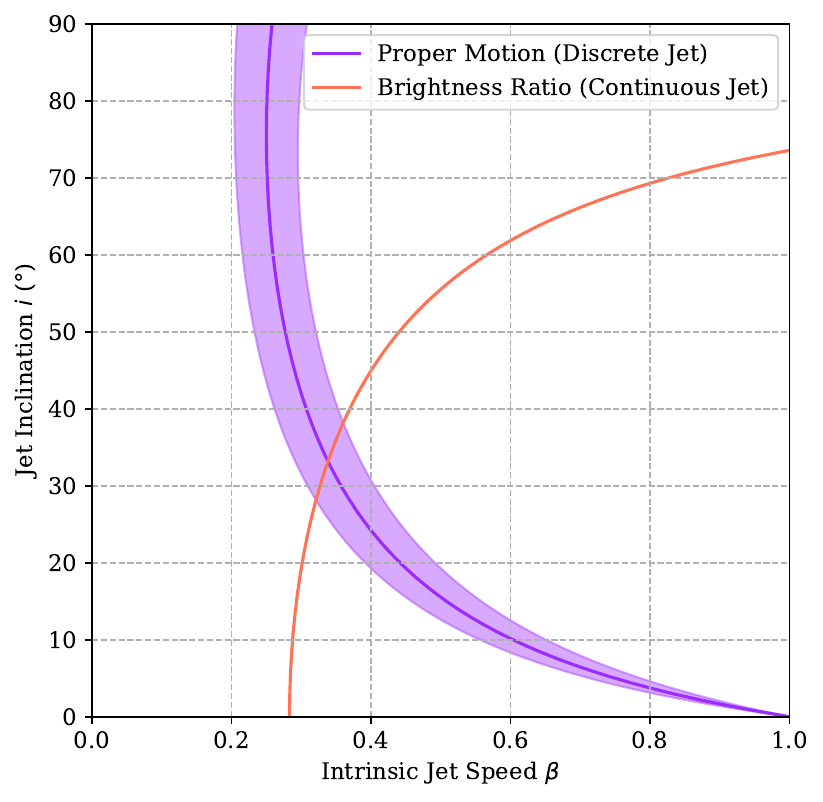}
        \caption{Constraints on the intrinsic speed and inclination of the extended and discrete jets seen in the VLBA observation in Figure~\ref{fig:VLBA Image 1}. The solid lines show the nominal values and the shaded areas show the $1\sigma$ uncertainties. We derive these constraints from the brightness ratio of the approaching and receding arms of the extended jet (equation~\ref{eqn:flux density ratio}) and the proper motion of the southern jet knot (equation~\ref{eqn:proper motion}). The brightness ratio puts an upper limit on the jet inclination and a lower limit on the intrinsic speed of the continuous jet.}
        \label{fig:jet speed inclination}
    \end{figure}

    \subsection{Southern Jet Knot}\label{sec:southern jet knot}
    While the nature of the jet knot is unclear, there are multiple potential scenarios. The most likely explanation is that the jet knot is the result of in-situ particle acceleration in the downstream continuous jet which produced synchrotron emission. Although the approaching extended jet expanded and faded beyond detection at a separation of $\sim30$ mas, the jet material will continue to propagate downstream until it loses its energy and momentum to the surrounding medium. An internal shock could be the result of the collision of fast-moving jet material with previously ejected slower-moving jet material. Internal shock models have been used in the past to explain both compact steady jets and discrete transient jets \citep{2010MNRAS.401..394J, 2014MNRAS.443..299M, 2018MNRAS.480.2054M}. A short lived internal shock could explain why the jet knot was fading rapidly and was not seen in subsequent observations. This model only requires a change in the jet speed of the continuous jet as it extends outwards during the rise of the outburst. In this scenario, the jet knot does not have to be travelling at the same speed as the continuous jet. 

    Alternatively, the particle acceleration could be due to a collision between the continuous jet and a dense interstellar medium (ISM). The discrete knot may have then been the `leading edge' of the extended jet as it advanced outwards through the ISM. Under this scenario, and assuming ballistic motion, the continuous jet would have began extending outwards from the core on MJD $60181.8\pm0.2$, which coincides with the peak of the initial hard-state X-ray rise. If the jet knot was decelerating, then the 'launch date' would have occurred later. In-situ particle acceleration due to the interaction between transient jets and the ISM has been observed to cause downstream rebrightening in multiple sources \citep[e.g.][]{2002Sci...298..196C, 2005ApJ...632..504C, 2017MNRAS.472..141M, 2020ApJ...895L..31E}, as well as downstream deceleration of the jet at both arc-second and mas scales \citep[e.g.][]{2002Sci...298..196C, 2010MNRAS.409L..64Y, 2011MNRAS.415..306M, 2019ApJ...883..198R, 2020ApJ...895L..31E, 2022MNRAS.511.4826C, 2023ApJ...948L...7B}, which may be the case here if the inclination of the jet axis is $\gtrsim37\degree$. This interaction between the continuous jet and the ISM would have required a dense local environment, which can be probed by observations of the motions of transient ejecta launched later in the outburst. This explanation does not account for the rapid fading of the jet knot during the VLBA observation, as the steady continuous jet should have continued to interact with the dense ISM across all of the observations. 

    A less likely explanation is that knot was a discrete transient jet, which travelled ballistically away from the core after being ejected on MJD $60181.8\pm0.2$ (or later if the knot was decelerating), similar to those that are often seen in other BH LMXBs \citep[e.g.][]{1994Natur.371...46M, 1995Natur.374..141T, 1995Natur.375..464H, 2012MNRAS.421..468M, 2020NatAs...4..697B, 2022MNRAS.511.4826C}. These transient jets fade and become optically thin as they expand, which could explain the fading of the jet knot observed here. In other X-ray binaries, these types of jets are ejected at the peak of the outburst, during the state transition \citep{2004MNRAS.355.1105F, 2009MNRAS.396.1370F}. In this scenario, this would be the earliest time in a BH LMXB outburst that a transient jet has been seen to have been ejected, occurring at the peak of the hard state (see Figure~\ref{fig:x-ray lightcurve}). Furthermore, these ejection events have previously been associated with specific X-ray signatures that correspond to changes in the inner accretion flow such as bright X-ray (and accompanying radio) flares, as well as a dramatic change in the X-ray spectral and timing properties \citep[e.g.][]{2009MNRAS.396.1370F, 2012MNRAS.421..468M, 2019ApJ...883..198R, 2020ApJ...891L..29H, 2021MNRAS.505.3393W}. \sjshort\, did exhibit many of these behaviours later in the outburst during the state transition, however, despite relatively intensive monitoring, no such clear signature of ejection was identified at the beginning of the outburst. An explanation for the jet knot that involves the re-acceleration of material in the out-flowing continuous jet does not require the ejection of a discrete transient jet knot during the hard/hard-intermediate state, and is more consistent with the current understanding of jet evolution during BH LMXB outbursts.

\section*{Acknowledgments}
We respectfully acknowledge the significant contributions made to this longstanding collaboration by Tomaso Belloni, who sadly passed away during our observing campaign. His insights and wealth of knowledge are sorely missed by his colleagues.

The National Radio Astronomy Observatory is a facility of the National Science Foundation operated under cooperative agreement by Associated Universities, Inc. This work made use of the Swinburne University of Technology software correlator, developed as part of the Australian Major National Research Facilities Programme and operated under licence. The Long Baseline Array is part of the Australia Telescope National Facility (\url{https://ror.org/05qajvd42}) which is funded by the Australian Government for operation as a National Facility managed by CSIRO. This work was supported by resources provided by the Pawsey Supercomputing Research Centre with funding from the Australian Government and the Government of Western Australia. This research has made use of the MAXI data provided by RIKEN, JAXA and the MAXI team. This work made use of the Warkworth 30m telescope as part of the LBA \citep{2015PASA...32...17W}. From 2023 July 1, operation of Warkworth was transferred from Auckland University of Technology (AUT) to Space Operations New Zealand Ltd, who continue to make the facilities available for VLBI out of goodwill. 

CMW acknowledges financial support from the Forrest Research Foundation Scholarship, the Jean-Pierre Macquart Scholarship, and the Australian Government Research Training Program Scholarship. FC acknowledges support from the Royal Society through the Newton International Fellowship programme (NIF/R1/211296). RF and SM acknowledge support from a European Research Council (ERC) Synergy Grant ``BlackHolistic" (grant No. 101071643). DMR is supported by Tamkeen under the NYU Abu Dhabi Research Institute grant CASS. AJT acknowledges the support of the Natural Sciences and Engineering Research Council of Canada (NSERC; funding reference number RGPIN-2024-04458). GRS is supported by NSERC Discovery Grant RGPIN-2021-0400. VT acknowledges support from the Romanian Ministry of Research, Innovation and Digitalization through the Romanian National Core Program LAPLAS VII – contract no. 30N/2023.
The authors wish to recognize and acknowledge the very significant cultural role and reverence that the summit of Maunakea has always had within the indigenous Hawaiian community. We are most fortunate to have the opportunity to conduct observations from this mountain. We also wish to acknowledge the Gomeroi, Gamilaroi, and Wiradjuri people as the traditional custodians of the LBA observatory sites. 


% \vspace{5mm}
\facilities{LBA, MAXI, NRAO, VLBA}


\software{AIPS \citep{1985daa..conf..195W, 2003ASSL..285..109G}, ADS (\url{https://ui.adsabs.harvard.edu/}), Arxiv (\url{https://astrogeo.org/}), Astrogeo (\url{https://astrogeo.org/}), Astropy \citep{astropy:2013, astropy:2018, astropy:2022}, CDS \citep[Simbad;][]{2000A&AS..143....9W}, Cmasher \citep{2020JOSS....5.2004V}, Corner \citep{corner}, Dynesty \citep{2020MNRAS.493.3132S}, eht-imaging \citep{2018ApJ...857...23C}, Jupyter \citep{Kluyver2016jupyter}, Matplotlib \citep{Hunter:2007}, Numpy \citep{harris2020array}, Scipy \citep{2020SciPy-NMeth}}




\bibliography{references}{}
\bibliographystyle{aasjournal}



\appendix
\section{Hard-state Jet Profile}\label{appendix:jet profile}
    To investigate the brightness ratio of the extended jet in Figure~\ref{fig:VLBA Image 1}, we extracted intensity profiles down the northern and southern jets. We first rotated the image of the jet so that it aligns vertically, and then took one dimensional cross sections of the northern and southern jets every 1.05 mas downstream from the core (7 pixels). We fit these cross sections with a one dimensional Gaussian profile to measure the intensity of the approaching and receding jets as a function of separation from the core, which we show in Figure~\ref{fig:jet profiles}. We fitted the Gaussian profiles to the pixel intensities using non-weighted least squares regression with \texttt{scipy.optimize.curve\_fit}\footnote{\url{https://docs.scipy.org/doc/scipy/reference/generated/scipy.optimize.curve_fit.html}}, where we report the intensity of the jet as the amplitude of the Gaussian profile. To calculate the uncertainty in the intensity, we added in quadrature the reported $1\sigma$ statistical uncertainty from the fit (extracted from the diagonal elements of the covariance matrix), the rms noise in the image, and a 10\% calibration error. In Figure~\ref{fig:jet profiles}, we also show the contribution of the unresolved core by taking the one dimensional profile of the restoring beam along the jet axis multiplied by the fit intensity of the core. The contribution from the unresolved core drops below the $5\sigma$ limit at a separation of $\sim3$ mas. We note that the measurements are correlated along the jet and across the core, since the true intensity profile of the jet is convolved with the Gaussian restoring beam. This explains why the decay of the northern jet is much steeper than the southern jet, since at smaller separations the convolution with the beam across the core causes the intensity to be biased upwards. 

    \begin{figure}[h]
        \centering
        \includegraphics[width=0.45\linewidth]{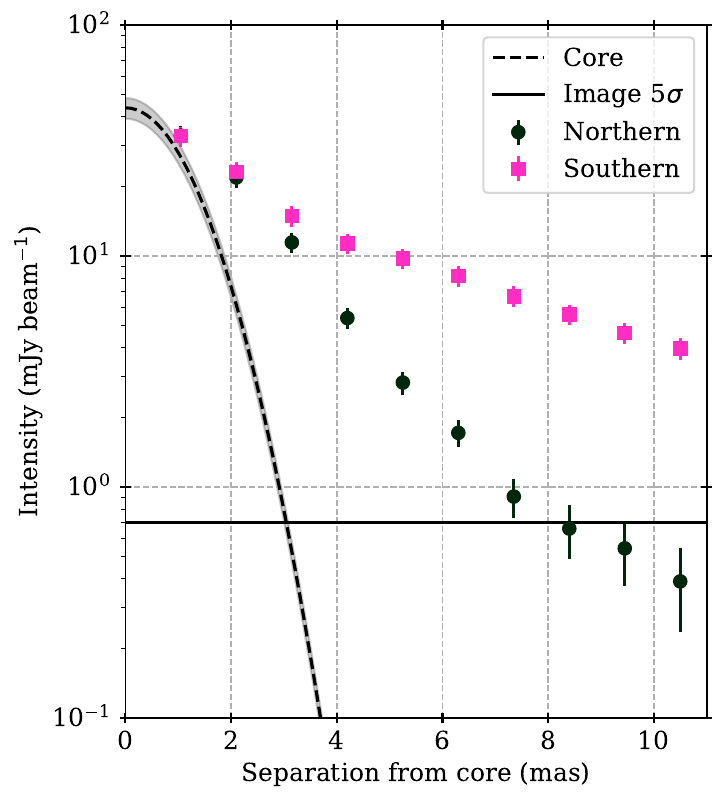}
        \caption{Intensity profile along the northern and southern jet as a function of separation from the core in the VLBA observation of \sjfull\; shown in Figure~\ref{fig:VLBA Image 1}. The intensity profiles are calculated by fitting 1D Gaussians perpendicular to the jet axis at intervals of $\sim1$ mas. The dashed line shows the contribution from the core, calculated by multiplying the intensity of the core by the profile of the restoring beam along the jet axis, with the shaded region showing its uncertainty (derived from the fit). }
        \label{fig:jet profiles}
    \end{figure}

% \allauthors
\end{document}